\documentclass{article}
 \usepackage{amssymb}

\author{G. Maiella  and C. Stornaiolo\\
	\\
}

\title{The Cardy-Verlinde equation in a spherical symmetric gravitational collapse}

\begin{document}

\maketitle

\begin{abstract}
 The Cardy-Verlinde formula is analyzed in the contest of the gravitational collapse. Starting from the holographic principle, we show how the equations for a homogeneous and isotropic gravitational collapse describe the formation of the black hole entropy. Some comments on the role of the entangled entropy and the connection with the c-theorem are made.
\end{abstract}

\section{Introduction}
In 1986 J. Cardy showed \cite{Cardy:1986ie},  using modular invariance,  that entropy of a 1+1 dimensional CFT is given by the expression
  \begin{equation}\label{1}
    S=2\pi\sqrt{ \frac{c}{6}\left(L_{0}-\frac{c}{24}\right)}.
\end{equation}
where  $L_{0}=ER$ and the term $\frac{c}{24}$  is the shift of energy due to the Casimir energy.

Aiming to extend this result to any dimension, Verlinde \cite{Verlinde:2000wg} noticed,  by using the holographic principle and  the ADS/CFT correspondence, that there is a striking similarity between the Cardy formula for entropy and the first of the FLRW  equations 
\begin{equation}\label{2}
     H^{2}=\frac{16\pi G}{n(n-1)}\frac{E}{V}-\frac{1}{R^{2}}
\end{equation}
\begin{equation}\label{3}
     \dot{H}=-\frac{8\pi G}{n-1}\left(\frac{E}{V}+p\right)+\frac{1}{R^{2}}
\end{equation}
of an  $(n+1)$ dimensional closed universe, where  $R$ is the factor expansion of the universe,  $H=\dot{R}/R$ is the Hubble constant and the mass-energy density is expressed by the ratio between energy and volume $E/V$.

The correspondence between equations (\ref{1}) and  (\ref{2})  is found   after making the following identifications
\begin{eqnarray}
       \label{4a}  2\pi L_{0} \Rightarrow  \frac{2\pi}{n} ER  \\
                        \label{4b}           2\pi \frac{c}{12} \Rightarrow  (n-1)\frac{V}{4GR}\\
             \label{4c}        S \Rightarrow  (n-1)\frac{HV}{4G},  
 \end{eqnarray}
this relation indicates the existence of a  general duality between quantum field theories and classical spacetime geometries at any dimension $D=n+1$.

Cardy's formula received considerable attention, because, as showed  in  \cite{Birmingham:2001qa} and in \cite{Medved:2004tp} an external observer finds out that the near-horizon region of a black hole is two dimensional and is conformal invariant, i.e. it is described by a rational CFT$ _{2}$. It has been  remarked by Carlip \cite{Carlip:2007qh}, that for this reason  Cardy's formula gives a clue for counting the statistical states that produce the macroscopic  area law for the entropy.

 Equations (\ref{2}) and (\ref{3})  describe also the gravitational collapse of a homogeneous spherically symmetric star \cite{Weinberg:gravandcosm}, in other words it describes  a black hole formation, this fact suggests a  close  correlation between the physics of black holes and conformal field theory.

 In  the  scheme proposed by Verlinde,  equation (\ref{3}) corresponds to the Euler equation with the contribution of  a non extensive  energy term like the  Casimir   energy
\begin{equation}\label{6}
     E_{C}=(E+pV-TS).
\end{equation}
Indeed, after making the follow identification
\begin{equation}
T_H=-\frac{\dot{H}}{2\pi H}
\end{equation}
 equation (\ref{3}) can be read as
\begin{equation}\label{5}
     E_{BH}=(E+pV-T_{H}S_{H})
\end{equation}
and $E_{BH}$ is the value of energy when $S_B=S_{BH}$. So  that in cosmology the Bekenstein-Hawking energy plays the r\^{o}le of the Casimir energy. The entropy bounds  imply a  lower bound $T\geq T_{H}$ for temperature \cite{Verlinde:2000wg}. 

In our previous work \cite{Maiella:2006hr} we discussed the properties of a BTZ black hole in ADS$_3$ in terms of an effective unitary CFT$_2$ with a central charge $c=1$ realized in terms of the Fubini-Veneziano vertex operators.  In this paper we deepen the study of the relationship between black hole and conformal theories by analyzing the evolution of the  entropy of a homogeneous perfect fluid  in a gravitational collapse according to  the holographic principle and  the Cardy-Verlinde equation. A statement of the holographic principle claims  that the entropy in a region  cannot  be larger than the Bekenstein-Hawking  area entropy of a black hole covering that region. So, as the collapse generates entropy,  the holographic bound can be saturated only at its end.  

Based on the principles set out and simple considerations one can show that the properties of a gravitational collapse are  not simply the result of evolution described by the equations (\ref{2}) and (\ref{3}),  but there are thorough physical implications.  First of all because the particles  are submitted to a strong  acceleration reaching  approximately the speed of light, so that they can be considered as  a fluid of radiation with equation of state to $p=\rho/n$  (in an $n+1$ spacetime) as their masses can be now neglected.  Moreover the entropy can grow monotonically and saturates the holographic limit only if the equation of state of the fluid takes the form $p=\rho$.  This fact suggests that the fluid undergoes an additional phase transition when the gravitational interaction becomes important.  

In section \ref{gravcoll}   we review the gravitational collapse of a homogeneous star in relation with  the Cardy--Verlinde equation, which shows how the total entropy of the star,  the Hubble entropy, is the sum of two terms in which the first term is  identified with the Bekenstein-Hawking area entropy and the second term in which the central charge of the underlying Conformal Field Theory (CFT)  is simply  proportional to the spatial curvature $k/R^2$ .  In section \ref{s3} we prove  that after the formation of a black hole,   there is a continuous formation of   concentrical inner black holes. The horizon of each of these black holes grows to coincide with the most external one once all the matter has fallen inside it.    According to the extended second principle of thermodynamics and the Cardy-Verlinde equation  this leads to an continuous increase of entropy,   and  Bekenstein-Hawking  bound  is saturated, when the collapse ends (or should we say that the collapse ends when the bound is saturated?). In section \ref{s4}   this result  is used to show that this process is equivalent to the evolution of the  the generic equation $P=(\gamma-1)\rho$  which brings us inevitably to $P=\rho$.

 We note that the results of the preceding paragraphs can be seen as the result of an underlying process of renormalization group  i.e. the evolution  equation of the adiabatic index $\gamma$ takes the form of a renormalization group equation. 
    We treated the gravitational collapse in a space with $3+1$ dimensions, but in section \ref{s5}  we show  that the results found can be extended to spaces of   $n+1$ dimensions with $n\ge 3$, some dimensional considerations are added. In section \ref{6} we end the paper discussing the results found and draw some conclusions.

\section{Gravitational collapse of a homogeneous star}\label{gravcoll}

 Verlinde's ansatz  generalizing   Cardy's  formula (from now on Cardy-Verlinde formula) lead to various attempts  aiming to verify the universality of CFT for the Black Hole physics and its cosmological cousin. The  closed FLRW universe model had a relevant role in extending   to arbitrary dimensions the formula for entropy because  two of the three terms in equation (\ref{2}) are proportional respectively to the Bekenstein, Bekenstein and Hawking entropy bounds, so one can assume that  the Hubble term is proportional to the total entropy of the system, but  this model describes the  spherically symmetric gravitational collapse  of a homogeneous star too. Indeed  from the equations \cite{Weinberg:gravandcosm}
\begin{equation}\label{7}
-2k-\ddot{R}(t)R(t)-2\dot{R}^{2}(t)=-4\pi G\varrho(0)R^{-1}(t)
\end{equation}
and
\begin{equation}\label{8}
   \ddot{R}(t)R(t)=-\frac{4\pi G}{3}\varrho(0)R^{-1}(t)
\end{equation}
 we can derive the  constraint  equation
\begin{equation}\label{9}
  \dot{R}^{2}(t)=-k+ \frac{8\pi G}{3}\varrho(0)R^{-1}(t).
\end{equation}
which coincides with equation (\ref{2}) in 3+1 dimensions.  (the entropy equation is independent from the equation of state of the material source).
Here we are considering a model with dust $P=0$. But  realistically  one should expect a change of equation of state of the initial dust fluid because during the collapse  the  particles forming the boundary of star are accelerated to $v\simeq c$. In other words when the particles approach the horizon they have  $p>>m_{0}$, i.e. the fields become "massless", so to  be described by a 
 conformally invariant field theory. 
 
 Let us  study the  "evolution" of entropy during in the collapse according to the holographic principle.

Eq. (\ref{9})   is derived    from the Cardy-Verlinde formula.  By slightly  modifying  the substitutions in (\ref{4a}), (\ref{4b}) (\ref{4c})  into   
\begin{eqnarray}
\label{a}2\pi L_{0} \Rightarrow  \frac{2\pi}{n\sqrt{k}} \frac{ER }{\ell_p^2}\equiv \frac{S_{B}}{\sqrt{k}} \\
 \label{b}                               2\pi \frac{c}{12} \Rightarrow    (n-1)\frac{\sqrt{k}V}{4GR\ell_p^2} \equiv (n-1)\sqrt{k} S_{BH} \\
\label{c}                           S \Rightarrow   (n-1)\frac{HV}{4G\ell_p^2} \equiv (n-1) S_{H} 
     \end{eqnarray}
where $\ell_p=(\hbar\, G/c^3)^{1/2}$ is the Planck length. The  relation
\begin{equation}\label{13}
   R_{S}=\frac{2GM}{c^{2}}= \frac{2G}{c^{2}}\frac{4\pi \varrho}{3}R^{3} = \frac{8\pi G}{3c^{4}}  \frac{E}{V} R^{3}
\end{equation}
is obtained from the definition of Schwarzschild radius and by multiplying and dividing it by the volume $V$ and will be useful in the following.
We find
\begin{equation}\label{14}
    \frac{V}{4GR}=\frac{1}{c^{4}}\frac{R}{R_{S}}\frac{2\pi}{3} ER
\end{equation}
Through this expression we find a relation between  $S_{B}$ given by Bekenstein (which differs of a factor $2$ from the one used by Verlinde) and $S_{BH}$ (using units $c=1$)
\begin{equation}\label{15}
 S_{BH}=\left( \frac{R}{R_{S}}\right)S_{B}
\end{equation}
i.e. both the entropy limits coincide when the radius of the collapsing sphere coincides the Schwarzschild radius.
This means that the collapsing system is in the weakly gravitating regime when
\begin{equation}\label{16}
    \frac{R}{R_{S}}>1
\end{equation}
and it is in a  strongly gravitating regime when
\begin{equation}\label{17}
    \frac{R}{R_{S}}< 1.
\end{equation}
By applying the substitutions (\ref{a}), (\ref{b})  and (\ref{c}) to the original Cardy-Verlinde formula and the definitions of entropy, we find
\begin{equation}\label{20}
  4S_{H}^{2}+\left(\frac{S_{B}}{\sqrt{k}}-2\sqrt{k}S_{BH}\right)^{2}=\frac{S^{2}_{B}}{k}
\end{equation}

These   relation  reduce to the following form
\begin{equation}\label{24}
   S_{H}^{2}= S_{BH}\left(S_{B}-k S_{BH}\right)
\end{equation}
where it is easy to prove that $k$, the curvature constant, is $k=R_{S} /R_{ 0 } $.
Then by using equation (\ref{15}), we find the following expression for entropy,

\begin{equation}\label{25}
    S_{H}^{2}=S_{BH}\left(S_{B}-S_{B}\frac{R}{R_{0}} \right)
\end{equation}
At the initial time $t=0$, when $R(t)/R_{0}=1$ so $R(t)/R_{0}<1$ for $t>0$,
\begin{equation}\label{26}
    S_{H}=0
\end{equation}
for subsequent times, it can be expressed in terms of, say, $S_{BH}$,
\begin{equation}\label{27}
  S_{H}(R)=S_{BH}(R)\sqrt{ R_{S}\left(\frac{1}{R}-\frac{1}{R_{0}}\right)}.
\end{equation}
and when  $R=R_{S}$
\begin{equation}\label{28}
     S_{H}(R_{S})=S_{BH}(R_{S})\sqrt{ 1-\frac{R_{S}}{R_{0}}}.
\end{equation}
This means that the Hubble entropy limit does not reach the  Bekenstein-Hawking black hole entropy at the black hole formation. It seems somewhat contradictory to the previous statements, but in  the next sections we prove that saturation is achieved only at the end of the collapse, when  the classical singularity is ``covered'' by ``quantum effects'' see e.g.\cite{Zurek:1984zz}\cite{'tHooft:1997py}.

\section{The gravitational collapse and the Cardy-Verlinde equation}\label{s3}

In this section we show how the  collapsing matter saturates the area bound at the end of  the gravitational collapse in a way consistent with the holographic principle by applying equation (\ref{28}). To this aim we
consider the initial spherical collapsing body with a homogeneous initial density $\rho_{0}$ and initial radius $R_{0}$ so that its mass is $M=4\pi \rho_{0} R_{0}{^{3}}/3$. Let us consider an infinite sequence of concentric spheres of this body,  labeled by an index $(n)$, with initial radius $R_{0}^{(n)}$ and with mass $M=4\pi \rho_{0} {R_{0}^{(n)}}{^{3}}/3$.
 
After the whole mass has formed the event horizon, the collapse continues under the Schwarzschild horizon by increasing the density $\rho$, and, by the relation between the Schwarzshild radius and the density at the time of the black hole formation (see \cite{Chandrasekhar:1964zza} \cite{Stornaiolo:2001nv}))
\begin{equation}\label{28a}
   M^{2}\propto
    R^{2}_{S}=\frac{3c^2}{8\pi G}\frac{1}{\rho},
\end{equation}
a smaller black hole with mass $M^{(1)}$ forms when the density takes the value
\begin{equation}\label{28b}
    \rho^{(1)}=\frac{3 c^6}{32 \pi G^{3} {M^{(1)}}^{2}}
\end{equation}
and, by  equation (\ref{28}), with entropy
\begin{equation}\label{28c}
     S_{H}(R^{(1)}_{S})=S_{BH}(R^{(1)}_{S})\sqrt{ 1-\frac{R^{(1)}_{S}}{R^{(1)}_{0}}}
\end{equation}
afterwards the black hole $(1)$ is accreted by the residual matter  its horizon grows to the Schwarzschild radius $R_{S}$,and finally the entropy of the total black hole with mass $M$ grows up to the value
\begin{equation}\label{28c}
     S_{H}(R_{S})\geq S_{BH}(R_{S})\sqrt{ 1-\frac{R^{(1)}_{S}}{R^{(1)}_{0}}}
\end{equation}
%
%
this process is iterated    until the singularity forms. At any increment of $n$ the final entropy grows. Indeed the ratio
\begin{equation}\label{28f}
    \frac{R^{(n)}_{S}}{R^{(n)}_{0}}= \left( \frac{\rho_{0}}{\rho^{(n)}}\right)^{1/3}
\end{equation}
where $\rho_{0}$ is the density of the spherical body at the starting time of the collapse.
Finally, taking the limit  $n\to \infty$ the density $ \rho^{(n)}\to \infty$ and 
\begin{equation}\label{28g}
     S_{H}(R_{S})=\lim_{n\to \infty}S_{BH}(R_{S})\sqrt{ 1-  \left(\frac{\rho^{\phantom{0}}_{0}}{\rho^{(n)}}\right)^{1/3}} =S_{BH}(R_{S})
\end{equation}
i.e. the Hubble entropy reaches the value of the area entropy.

For sake of clearness this process has been described as a discrete process, but it can be thought as   continuous. In conclusion we have shown that as a consequence of the holographic principle and of the Cardy-Verlinde equation the  Hubble entropy $S_{H}(R_{S})$ coincides  with the Bekenstein-Hawking entropy $S_{BH}$, then there is  a connection between the entropy seen by an external observer and the entropy formed inside a black hole.

We stress that the construction  introduced here recalls the one  used by \cite{Balasubramanian:1999jd} in association with a renormalization group equation. Actually,  in the next section we prove that   the  saturation of the holographic bound follows from  the evolution of the equation of state  described by a  renormalization group equation.  Our aim is to show that these results have very far reaching consequences. In such a context we wish to analyze briefly the relation between entropy, c-heorem and entanglement.  Let us start with  a brief analysis of the c-theorem in 2d  CFT$_2$ by stressing its relation with the black hole entropy and to the generalized second principle of thermodynamics.

In references \cite{Zamolodchikov:1986gt} \cite{Casini:2006es} \cite{Casini:2004bw} the c-theorem in CFT$_2$ was shown to hold.    Its formulation of interest here is   the following:

For any QFT in two euclidean dimensions there is a  dimensionless universal function of the distance $c(R)$,  where  $R=x_1^2 + x_2^2$,  which is  non increasing under dilatations $x_i\to \lambda x_i$
and takes a finite value proportional to the central charge $c$  at the fixed point. The original proof requires that the energy-momentum tensor $T_{ij}$ has to be unitary. It should be noticed that in the proof given in \cite{Casini:2006es}\cite{Casini:2004bw} \textit{ the relation of the c-function with the entanglement entropy (and the information theory) is  made explicit. }

We  stress that the relation between the renormalization group, the c-theorem, the entanglement entropy  and the generalized second principle of thermodynamics is quite explicit  in the case  of the black hole collapse considered here and in a previous work on ADS$_3$\cite{Maiella:2006hr}\cite{Maiella:2008jn}. Naturally some points need  to be further clarified, but once again the ''imprinting'' of QFT near the IR critical point (i.e. CFT) seems to be robust at least for { \it{any even dimensions}} $n$.

\section{Phase transitions inside a black hole}\label{s4}

 According to the holographic principle  the maximal entropy in a region (for a strongly gravitating system) is given by  the area entropy of a black hole contained in such a region. In the previous section we have seen how the application of  the  holographic principle implies that the entropy of the collapsing matter saturates this entropy bound.  This result has an interesting consequence on the behavior of the collapsing matter.

As  equation (\ref{24}) is obtained by multiplying both sides of equation (\ref{9})
by
\begin{equation}\label{28-1}
    \frac{3V^{2}}{16\pi G R^{2}\ell_p^4}
\end{equation}
 it can be written also as
\begin{equation}\label{9b}
 S_H^{2}\equiv  \frac{3V^{2}H^{2}(t)}{16\pi G\ell_p^4}=-\frac{3kV^{2}}{16\pi G R^{2} \ell_p^4}+  \frac{\varrho V^{2}}{2\ell_p^4}.
\end{equation}
and if  $P=0$ the right-hand side of the equation  goes to zero for $V\to 0$, leading to an inconsitency  with the  results found in the last section.

It follows  that the equation of state of the fluid has to change during the collapse.


It is well-known that equations (\ref{8}) and (\ref{9}) imply the conservation equation
\begin{equation}\label{50}
    \dot{\rho}=-3\frac{\dot{R}}{R}\left(\rho+P\right).
\end{equation}
This equation  implies a general relation between $P$ and $\rho$ \cite{Stornaiolo:1994mw},
\begin{equation}\label{51}
   P=\left(-\frac{1}{3}\frac{R}{\rho}\frac{d\rho}{d R}-1\right)\rho
\end{equation}
with
\begin{equation}\label{52}
  \gamma= - \frac{1}{3}\frac{R}{\rho}\frac{d\rho}{d R}
\end{equation}
 function of $\rho$ or of $R$.
 
 $\gamma$ is the adiabatic index given by the ratio of  enthalpy and internal energy and therefore by the ratio betweem
 The heat capacities at constant pressure $C_p$ and constant volume $C_V$
 \begin{equation}
\label{2051}
\gamma=\frac{C_P}{C_V}
\end{equation}
 
 To interpret $\gamma-1$ we write the perfect gas equation of state in the form
  \begin{equation}
\label{2052 }
P=\rho_m RT
\end{equation}
where $R$ is a the gas constant, $T$ the temperature and $\rho_m$ is the mass density, so that 
\begin{equation}
\label{2053}
C=\sqrt{RT}
\end{equation}
is the characteristic thermal speed of the "molecules". Then
\begin{equation}
\label{2054 }
\gamma-1=\frac{\rho_m C^2}{\rho_m c^2}=\frac{C^2}{ c^2}
\end{equation}
gives the rate between the square of the  thermal speed and the square of the light speed.

If  $\gamma=const.$, $C$ coincides with  the speed of sound
\begin{equation}\label{53}
    \frac{c^{2}_{s}}{c^{2}}=\frac{dP}{d\rho}
\end{equation}
and $\gamma$  can admit only values running from $1$ to $2$ so that the speed of sound is real and less or equal than the speed of light $c$.


In  general  the speed of sound may change   satisfying the equation
\begin{equation}\label{54}
    \frac{c^{2}_{s}}{c^{2}}=\frac{dP}{d\rho}=\rho\frac{d\gamma}{d\rho}+\gamma(\rho)-1
\end{equation}
or if we take $\gamma=\gamma(R)$\footnote{Equation (\ref{55}) is a Bernoulli equation and can be easily solved by assigning an evolution for $c_{s}^{2}$, its solution is
$$
\gamma(R)=\frac{\exp\left(-3\int_{R_{0}}^{R}dR'F(R')/R'\right)}
{-3\int_{R_{0}}^{R}1/R'\exp\left(-3\int_{R_{0}}^{R'}dR''F(R'')/R'' \right)dR'}
$$
where $F(R)=c^{2}_{s}(R)+1$.}
\begin{equation}\label{55}
    \frac{c^{2}_{s}}{c^{2}}=\frac{dP}{d\rho}=-\frac{1}{3}\frac{R}{\gamma}\frac{d\gamma}{dR}+\gamma(R)-1
\end{equation}

Both  equations (\ref{54}) and (\ref{55}) take the form of  renormalization group equations  for the adiabatic index  during the gravitational collapse.  

From a qualitative point of view
 the constituents  of the fluid accelerate and when  their velocities approach  the speed of light,  their rest  mass energy can be neglected with respect to their kinetic energy. At this stage the fluid behaves  as  radiation  with $\gamma=4/3$.  But  a  further increase  to $\gamma=2$   such that
\begin{equation}\label{29}
   P=\rho
\end{equation}
 is still possible when the gravitational attraction becomes important in the interactions internal to the fluid.
In this case the energy density scales as 
\begin{equation}\label{29bis}
     \rho=\mathcal{ A} R^{-6}
\end{equation}
where
\begin{equation}
\label{1000}
\mathcal{A}=18\frac{G^4 M^4}{c^8 }
\end{equation}
 is a  constant  with dimension of a mass times a volume fixed  requiring that,  when   $R\rightarrow 0$,  , is saturated by the  Hubble entropy $S_H$  is equal to the Bekenstein--Hawking area law according to eq.(\ref{9b}).

Stiff matter properties  have been widely studied.  In   \cite{Fischler:1998st}  \cite{Banks:2001yp}  it has been proved that in any FLRW with a homogeneous fluid and equation of state $p=\rho$ the entropy contained within a sphere of radius equal to the particle horizon scales with area of the sphere. 

 Considering a self-gravitating fluid with a central black hole  \cite{Zurek:1984zz} and  when the  Hawking radiation produced by a black hole  is treated  as a self-screening atmosphere \cite{'tHooft:1997py}, it was proved that the ratio between the entropy $S$ and the area $\Sigma$ is for any equation of state $P=(\gamma-1)\rho$
\begin{equation}
\label{2008 }
\frac{S}{\Sigma}=\frac{\gamma}{7\,\gamma-6}
\end{equation}
  it comes out that the stiff matter equation of state has the  interesting property that i.e. this ratio is equal to $1/4$ when $\gamma=2$. 

%

 The equation of state $P=\rho$ has  been also deduced in \cite{Diaz:2005ga} for a fluid of black holes. In \cite{Zurek:1984zz} \cite{'tHooft:1997py} and \cite{Banks:2001yp} the statistical mechanics of the $P=(\gamma-1)\rho$ was studied,  and the  temperature scaling of  entropy and energy densities is
\begin{equation}
\label{1001 }
s\sim T^{\frac{1}{\gamma-1}}
\end{equation}
and
\begin{equation}
\label{1002 }
\rho\sim T^\frac{\gamma}{\gamma-1}
\end{equation}
It follows that the thermodynamics  a stiff matter fluid corresponds to  the thermodynamics of a conformal theory in $1+1$ dimensions 
\begin{equation}
\label{1003 }
s\sim T
\end{equation}
and
\begin{equation}
\label{1004 }
\rho\sim T^2.
\end{equation}
An example of stiff matter is given by a non elementary scalar field $\phi$  when its potential  $ V(\phi)$ goes to zero. The components of the scalar field  energy-momentum tensor,  interpreted as the components of an fluid energy momentum tensor are
\begin{equation}\label{300}
    \rho=\frac{1}{2}\dot{\phi}^{2}+ V(\phi)
\end{equation}
and
\begin{equation}\label{310}
    p=\frac{1}{2}\dot{\phi}^{2}- V(\phi)
\end{equation}
 where $\phi=\phi(t)$  When $ V(\phi)=0$, we have $P=\rho$ equation of state. .The conservation equation corresponds  to the Klein-Gordon equation,
\begin{equation}\label{320}
     \ddot{\phi}+3H\dot{\phi}+ V'(\phi)=0.
\end{equation}
%

It is possible to show that to any given an equation of state it corresponds a family of potentials $V(\phi)$ \cite{Stornaiolo:1994mw}. For the potential
\begin{equation}\label{340}
     V(\phi)=\frac{2-\gamma}{8}B\left(\exp(-\sqrt{\alpha}\phi)-2D+D^{2}\exp(\sqrt{\alpha}\phi)\right).
\end{equation}
with a constant  $\gamma$ is equivalent to he  equation of state is $p=(\gamma-1)\rho$,  between (\ref{300}) and (\ref{310}), where  $B$ and $D$ are arbitrary constants and $\alpha=24\pi G \gamma$ (see ref. \cite{Stornaiolo:1994mw} for details). From equations (\ref{320}) and (\ref{340}), it follows that the relation between $\rho$ and $\phi$ is
\begin{equation}\label{350}
     \phi=\frac{1}{\sqrt{\alpha}}\ln\left[ \frac{2\rho}{B}+D+2\sqrt{\frac{\rho^{2}}{B^{2}}+\frac{D\rho}{B}}\right]-\frac{2}{\sqrt{\alpha}}\ln D.
\end{equation}
The potential vanishes   when  $\gamma\to 2$,  which  corresponds to a sort of asymptotic freedom between the elementary constituents of the fluid.

\section{Some dimensional considerations}\label{s5}

So far we have considered the usual gravitational collapse in $3+1$ dimensions, but it is easy to show that all the results obtained in the previous sections can be directly extended to any $n+1$ spacetime with  eq. (\ref{2})
\begin{equation}\label{2001}
     H^{2}=\frac{16\pi G}{n(n-1)}\frac{E}{V}-\frac{k}{R^{2}}
\end{equation}
and the conservation equation 
\begin{equation}\label{2002}
\dot{\rho}+n\frac{\dot{R}}{R}(\rho+P)=0
\end{equation}
where $\rho$
 and $P$ are  the mass-energy density and the pressure for a fluid in $n+1$ dimensions with energy-momentum tensor
 \begin{equation}\label{2003}
T_{ab}= Diag(\rho,\underbrace{ -P,....,-P}_{n\rm{\  times}})
\end{equation}
and we note that imposing the trace of the energy-momentum tensor for radiation i equal to zero implies that
\begin{equation}\label{2004}
P=\frac{1}{n}\rho.
\end{equation}
From (\ref{2002})  it follows  that when $P=\rho$,
\begin{equation}\label{2005}
\rho\propto \frac{1}{R^{2n}}
\end{equation}
i.e. the energy density  scales as the square of a spatial volume, so that,  according to the correspondences introduced by Verlinde's  (see eqs. (\ref{4a})  (\ref{4b}), (\ref{4c}) and (\ref{28-1})),  in any dimension the entropy  goes to a finite value at the end of the collapse saturating  the Bekenstein-Hawking bound.

Let us  remark that for $n=1$ in equation  (\ref{2004}), the null trace condition  for radiation coincides with the equation of state $P=\rho$, from which we can derive all the thermodynamical properties of radiation in a $1+1$ spacetime.
It is interesting to note that this condition and equation (\ref{2002})  imply  in $1+1$  that 
\begin{equation}\label{2006} 
\rho\propto \frac{1}{R^2}
\end{equation}
which  for  $n\ge 3$  is the relation between $\rho$ and the Schwarzschild radius given by eq.(\ref{28a}), indeed in \cite{Diaz:2005ga}  it was shown that at any dimension the number density 
\begin{equation}\label{2007}
\mathcal{N}\sim R_s^{n}
\end{equation}
  combined with the relation for the mass $M\sim R_s^{n-2}$ gives  eq.(\ref{28a}) for any $n$.

\section{Discussion and conclusions}\label{6}

 Verlinde's generalization of the Cardy formula to  dimensions higher  than two is based on the Friedmann equations for a closed universe. In this paper we point out  that the reason why this extension works is that  these equations  are related to the black hole formation as they describe the gravitational collapse of a homogeneous spherical star. 
  
Consistently with the holographic principle   the area entropy is saturated only at the end of the collapse, i.e. when  the classical singularity forms. According to equation (\ref{9b}),  the entropy of collapsing fluid  is equal to  the Bekenstein-Hawking entropy   only if  the final equation of state is $P=\rho$ (for all the other equations of state it  goes to zero  with  the volume).   It follows that at the last stages of the collapse  the thermodynamical properties  of the fluid  are  described by   a CFT$_2$. This result is  in agreement with the results found in literature \cite{'tHooft:1997py}\cite{Fischler:1998st}\cite{Banks:2001yp}\cite{Diaz:2005ga}.  We showed that   the analysis in section \ref{s3}, can be extended to any dimension $n$,  this implies that  the quantum processes at very high densities can be treated with the methods of  conformal field theory at any dimension. We found that the quantum effects   implicit in eq. (\ref{1}) (and exact in $2D$) "cover" the classical singularity,  because  the generation of an  infinite sequence of  internal concentrical  black holes   lead to a  production of  Hawking radiation at an increasing rate so that the collapse  is  accompanied by a phase  transition (of quantum origin) of  matter into radiation. 

 In  the papers of Zurek and Page, and  't Hooft  a generalization  of the Tolman-Oppenheimer-Volkoff equation was considered  for a self gravitating fluid composed by the Hawking particles  containing a  central black hole. The radiation  ''feels'' a negative point mass at $r=0$.   Here we find a similar configuration  in  a dynamical evolution and   out of  thermodynamic equilibrium. These results suggest that the equation of state for stiff matter, may have  physical meaning at very high densities and at very high energies, very close to the classical singularity. On the other side, the sequential formation of smaller horizons  during the collapse introduced in sect.  \ref{gravcoll}, leads to an increasing formation of  Hawklng radiation inside the black hole. So there must be a configuration Hawking radiation-black hole similar  to the one set in
\cite{'tHooft:1997py} except for the fact that here we do not have thermal equilibrium but a dynamic evolution of this system. 
In such a context we notice a quite original similarity with the Quantum Hall  Effect (QHE) \cite{Cristofano:1993nb}. In fact the QHE at Laughlin fillings $v=1/(2p+1) $ can be described by a CFT$_2$
with central charge $c=1$. It has been  noted in  \cite{Cristofano:1993nb} that the boundary (Casimir)  effect induces the correct electric force of sign opposite to the repulsive force between electrons. That results on the ''confinement" 
of them inside the cylinder (no spilling out!). Naturally the trace anomaly of the tensor $T_{\mu\nu} $ there is equal to $c/6$ (as for the gravity) which gives raising to the Casimir energy and for the Black Hole in ADS$_3$ to the dissipative Hawking radiation \cite{Maiella:2006hr}.
 Obviously it deserves a further investigation which we postpone to  forthcoming paper.

For the moment we observe that  the presence of Hawking radiation inside the black hole may be also interpreted as a change of vacuum where all  the microphysical processes  involved should therefore be described  by the quantum field theories  at finite temperature. The microstates of the black hole for  $P=\rho$  have a thermodynamical behavior identical to the one of CFT in two dimensions (see section \ref{s4}). 

Then the  formation and subsequent vanishing, by emission of Hawking radiation, of the Black Hole is a dynamical quantum process. To us it looks very similar to the renormalization group in QFT. In fact the collapse, classically would never end (i.e. at the centre of the ''sphere'' one finds a ''singularity'' but the quantum fluctuations of the  vacuum become stronger and stronger when the system approaches the singularity. In a pictorial way we can say that the fluctuations produce small black holes which screen the ''would be'' singularity at the center of the sphere. One would recognize a similarity with the ''atmosphere'' introduced in  \cite{Zurek:1984zz}\cite{'tHooft:1997py}. Both pictures remind us also of the methods of ''imaging'' used to find the solutions of electromagnetic static potentials. 

Finally it must be noticed that also in this work the thermodynamic properties seem to be a ''dual'' description of gravity \cite{Jacobson:1995ab} \cite{Padmanabhan:2009vy} \cite{Verlinde:2010hp}.  

In conclusion the use of quantum field theory near the IR point (i.e. CFT) encompasses all the just presented physical properties of the process giving   us a quite clear link between the c-theorem of CFT in any dimension with the dissipative system called Black Hole.

Acknowledgements We thank Prof. P. Santorelli (University "Federico II" of Naples) and Dr. C. Bachas (ENS-Paris) for an interesting discussion on the c-theorem for $D>2$ and for encouraging and useful suggestions.


\end{document}